\documentclass[aps,floatfix]{revtex4}

\usepackage{graphicx}
\usepackage{epsfig}
\usepackage{amssymb}

\def\lsim{\mathrel{\rlap{\lower4pt\hbox{\hskip1pt$\sim$}}
    \raise1pt\hbox{$<$}}}         
\def\gsim{\mathrel{\rlap{\lower4pt\hbox{\hskip1pt$\sim$}}
    \raise1pt\hbox{$>$}}}         

\begin{document}




\title{
Quark Coalescence based on a Transport Equation}
\author{L. Ravagli and R. Rapp}
\affiliation {Cyclotron Institute and Physics Department, Texas A\&M University, College Station, Texas 77843-3366, U.S.A.}
\date{\today}

\begin{abstract}
We employ the Boltzmann equation for describing hadron production from a 
quark-gluon plasma (QGP) in ultrarelativistic heavy-ion collisions.  
We propose resonance formation in quark-antiquark scattering as the 
dominant meson-production channel, which, in particular, ensures that
energy is conserved in the recombination process. This, in turn, 
facilitates a more controlled extension of ha\-dronization to low
transverse momenta ($p_T$), and to address the experimentally observed 
transition from a hydrodynamic regime to constituent quark-number scaling 
(CQNS).
Based on input dis\-tributions for strange and charm quarks with 
azimuthal asymmetries, $v_2(p_T)$, characteristic for RHIC energies, we 
recover CQNS at sufficiently high $p_T$, while at low $p_T$ a scaling
with transverse kinetic energy is found, reminiscent to experiment.
The dependence of the transition regime on microscopic QGP properties, 
i.e.~resonance widths and $Q$-values in the 
$q+\bar q \to M$ process, is elucidated. 
\end{abstract}

\maketitle

\section{Introduction}
\label{sec_intro}
Among the surprising experimental results in Au-Au collisions in the 
first years of operation of the Relativistic Heavy-Ion Collider 
(RHIC) were several unexpected features of hadron spectra in the 
intermediate transverse-momentum regime, 
$p_T\simeq2-6$~GeV~\cite{Adcox:2004mh,Adams:2005dq}. 
Most notably, a large baryon-to-meson ratio of around one and the 
so-called constituent quark-number scaling
(CQNS) of the elliptic flow coefficient, $v_2(p_T)$, have been found. 
These features are difficult to explain in hydrodynamic 
models~\cite{Teaney:2001av,Hirano:2002ds,Kolb:2003dz} which 
provide an excellent description of hadronic observables in the 
low-$p_T$ region, $p_T\le 2$~GeV, but lack sufficient yield 
at intermediate $p_T$.   
Quark coalescence models (QCMs) have been used extensively and 
successfully to describe the baryon-over-meson enhancement and CQNS at 
intermediate 
$p_T$~\cite{Hwa:2002tu,Greco:2003mm,Fries:2003kq,Molnar:2003ff,Pratt:2004zq}. 
The basic mechanism underlying QCMs is the recombination of constituent
quarks into hadrons at the putative phase boundary 
(gluonic degrees of freedom are either neglected or converted ``by hand" 
into $q$-$\bar q$ pairs). The constituent quarks are identified with the 
{\it valence} quarks inside hadrons, and, consequently, the hadron 
spectra directly reflect upon the partonic spectra and collective 
dynamics of the Quark-Gluon Plasma (QGP). A major challenge 
is then to search for (and possibly construct) a unified description of 
hadron production at low and intermediate $p_T$. In fact, as 
demonstrated in a recent PHENIX paper~\cite{Adare:2006ti}, a very 
general CQNS seems to emerge (in centrality, system size and collision
energy) when the scaling is applied to the {\em kinetic energy} of the 
produced hadrons.      


One of the limitations of conventional QCMs is the collinear and instantaneous 
approximation of the hadron formation process which limits their 
applicability to intermediate $p_T$ where the inherent violation of 
energy conservation is expected to be
small. It is also not obvious how to recover an equilibrium limit
in this framework which renders the connection to the low $p_T$ 
(hydrodynamic) regime less transparent.   
In this work, we propose an alternative realization of the recombination 
picture by formulating hadron production in terms of a Boltzmann transport 
equation.  Our key dynamical ingredient is hadronic resonance formation via 
$q$-$\bar q$ ``annihilation". Their finite width not only enables to go 
beyond the collinear 
limit and explicitly conserve 4-momentum, but also introduces, in principle,
a sensitivity to spectral properties of the QGP (e.g., widths 
and $Q$ values in the fusion process).   
In the limit of large times, on which we will focus here, we recover an 
equilibrium formula which reproduces the thermal Boltzmann limit and can 
be compared to QCM results. We will constrain ourselves to
the cases of charm and strange quarks hadronizing within a bulk medium
where the input anti-/quark phase distributions are parametrized 
with collective properties inferred from RHIC data. We will compute both
$p_T$ spectra and $v_2(p_T)$ for $\phi$, $D$ and $J/\psi$ mesons, and study
their scaling properties in connection with the transition from low 
to intermediate $p_T$.    
Our approach bears some similarity with recent work in 
Ref.~\cite{Miao:2007cm}, which, however, does not address the $v_2$ problem 
and utilizes different interactions.

\section{Boltzmann Equation with Quark-Antiquark Resonances}
\label{sec_boltz}
Our starting point is the Boltzmann equation for the meson phase-space
distribution, $f_M$, 
\begin{equation}
\label{boltzmann}
(\frac{\partial f_M}{\partial t} +\vec{v}\cdot\vec{\nabla}{f_M})(x,p)
=-\Gamma/\gamma_p \ f_M(x,p)+\beta(x,p) \ , 
\end{equation}
where $p$ and $x$ denote 3-momentum and position of the meson $M$, and  
$\Gamma$ its width which in the present work is attributed
to 2-body decays into quark and antiquark, $M\to q + \bar q$, and
assumed to be constant.    
The factor $\gamma_p=E_M(p)/m$ ($m$, $E_M(p)$: meson mass and energy), 
is due to Lorentz time dilation (see also Ref.~\cite{Miao:2007cm}). 
The surface term on the left-hand-side of Eq.~(\ref{boltzmann}) 
is mostly relevant for high-$p_T$ particles, and vanishes in the 
infinite volume limit; since our objective here is not a quantitative
description of data we neglect it in the following. 
The relation of the gain term, $\beta(x,p)$, to the underlying microscopic
interaction becomes explicit upon integration over phase space,  
\begin{equation}
\label{gain}
B \equiv \int \frac{d^3~p~d^3~x}{(2\pi)^3}~\beta(p,x)
= \int~\frac{d^3~x~d^3~p_1~d^3~p_2}{(2\pi)^6}~ f_q(x,p_1) 
f_{\bar{q}}(x,p_2)~\sigma(s)~v_{rel}(p_1,p_2)
\end{equation}
with $\sigma(s)$ the cross section for the process $q+\bar{q}\to M$ at 
center-of-mass (CM) energy squared $s=(p_1^{(4)}+p_2^{(4)})^2$,  
$p_{1,2}^{(4)}$ the 4-momenta of quark and antiquark, and 
$f_{q,\bar{q}}$ their phase space distribution functions, normalized as 
$N_{q\bar{q}}=\int \frac{d^3 x~d^3 p}{(2\pi)^3} f_{q,\bar{q}}(x,p)$. 
Throughout this paper, quarks will be assumed to be zero-width quasi-particles 
with an effective mass $m_q$ which we treat as a parameter. 
The intrinsically classical nature of the Boltzmann equation warrants the
use of classical distribution functions for all the particles. In addition, 
we will assume zero chemical potential for all quark species.
For the cross section we employ a relativistic Breit-Wigner form,  
\begin{equation}
\label{cross}
\sigma(s)=g_{\sigma}\frac{4\pi}{k^2}
\frac{(\Gamma m)^2}{(s-m^2)^2+(\Gamma m)^2} \ ,
\end{equation}
where $g_{\sigma}=g_M/(g_q g_{\bar{q}})$ is a statistical weight given in 
terms of the spin (-color) degeneracy, $g_M$ ($g_{q,\bar{q}}$), of the 
meson (anti-/quark); $k$ denotes the quark 3-momentum in the CM 
frame.  With $M\leftrightharpoons q +\bar q$ being the only channel, it 
follows that $\Gamma_{in}=\Gamma_{out}=\Gamma$. Detailed 
balance requires the same $\Gamma$ in the loss term on 
the right-hand-side of Eq.~(\ref{boltzmann}),
thus ensuring a proper equilibrium limit with $\tau=1/\Gamma$ the pertinent
relaxation time. This formulation conserves 4-momentum
and applies to all resonances $M$ with masses close to or
above the $q\bar q$ threshold, i.e., for positive $Q$ value,  
\begin{equation}
\label{masscondition}
Q= m - (m_q+m_{\bar q}) \gsim 0 . 
\end{equation}
If the $2\rightarrow 1$ channel proceeds too far off-shell, i.e., $Q<0$ and 
$\Gamma < |Q|$ (e.g., for pions), other processes need to be considered, 
e.g., $q+\bar{q}\rightarrow M+g$ (which is possible  
in the present formalism by including the respective cross sections). 
   


Let us now elaborate the equilibrium limit of our approach, by imposing the
stationarity condition, 
\begin{equation}
\label{eqcond}
0 = \frac{dN_M}{dt}|_{eq} = 
-\int \frac{d^3x~d^3p}{(2\pi)^3}\Gamma f_M^{eq}(p)/\gamma_p + B . 
\end{equation}
Introducing the notation $p_M=p_1+p_2, p_{rel}=p_1-p_2$ into the gain term 
$B$, eq.~(\ref{gain}), we find  
\begin{eqnarray}
\label{eqboltzappr}
N_{M}^{eq}&=&\int\frac{d^3x ~d^3p_M}{(2\pi)^3}f_{M}^{eq}(p_M,x) \ ; \quad 
f_M^{eq}(p,x)\equiv g(p_M,x)\gamma_p/\Gamma
\\
\label{gainterm}
g(p_M,x) &\equiv & \int\frac{d^3p_{rel}}{8~(2\pi)^3} 
\ f_q(x,p_M,p_{rel}) \ 
f_{\bar{q}}(x,p_M,p_{rel}) \ \sigma(s)~v_{rel}(p_M,p_{rel}) \ . 
\end{eqnarray}
Eq.~(\ref{eqboltzappr}) represents the large time limit of the Boltzmann 
equation and is the expression which comes closest to the conventional 
QCM formula. 
For hadronization times less or comparable to the relaxation time, $\tau$,  
the equilibrium limit will not be reached and a short time solution will 
be appropriate. In that case, the time variable enters explicitly 
into the final result, reflecting the dynamical 
nature of the Boltzmann equation.

We have verified numerically that for $\Gamma\to 0$ 
Eq.~(\ref{eqboltzappr}) accurately recovers the standard Boltzmann 
distribution for a meson $M$ at temperature $T$, {\em if} the constraint
of a positive $Q$ value is satisfied (for negative $Q$ 
the 2$\to$1 channel is inoperative).    
This shows that equilibration and energy conservation are closely related 
in our approach, constituting a significant improvement over previous QCMs.


\section{QGP Fireball, Spectral Properties and Transverse-Momentum Spectra}
\label{sec_fireball}
To compute meson spectra we have to specify the input $q$ and $\bar q$ 
distributions (including their masses and collective properties), 
as well as the meson-resonance masses and widths. 
For an exploratory calculation, we will focus on a QGP fireball close to 
the expected hadronization temperature, $T_c\simeq170$~MeV (as in previous
QCM studies). 
For simplicity, we will assume a constant (average) homogeneous cylindrical 
volume with collective transverse expansion velocity with linear radial 
profile, 
${\vec v}_T(r_T)=\beta_0\frac{{\vec r}_T}{R_T}$ ($R_T$, $\beta_0$: surface
radius and flow velocity), and use the quasi-equilibrium limit, 
Eq.~(\ref{eqboltzappr}). 
The quark distributions in the local (thermal) frame take the form
\begin{equation}
\label{fq}
f_q(p,x)=\exp\left[-\gamma_T (E_q-\vec{p}\cdot\vec{v}_T)/T\right]
\end{equation}
where $\gamma_T=(1-v_T^2)^{-1/2}$ and $E_q=(m_q^2+p^2)^{1/2}$ is
the quark energy in the lab frame.   
In this section we evaluate $p_T$ spectra for central 
$\sqrt{s_{NN}}=200$~GeV Au-Au collisions, thus neglecting any azimuthal 
asymmetries. While the shape of the $p_T$ spectra is affected by 
$\beta_0$, the total multiplicities must be independent of $\beta_0$.
Based on the fireball eigenvolume, $V^*=\int d^3x^*$, corresponding to
a system at rest ($\beta_0=0$), the volume element in the lab frame 
follows from the corresponding Lorentz-contraction factor
$d^3x_{lab}=d^3x^*/\gamma(x)$. We have verified that this leads to 
total quark multiplicities independent of $\beta_0$. 
The eigenvolume is fixed at $V^*\simeq1500$fm$^3$ using a hadron resonance 
gas in chemical 
equilibrium at $T\simeq170$~MeV to reproduce the correct pion rapidity 
density $dN_\pi/dy(y=0)$ for 0-10\% central collisions (including 
feed-down)~\cite{Rapp:2000pe}. 
Since thermal rapidity distributions for heavy particles are 
narrower than for thermal pions, the $dN/dy(y=0)$ values 
for the former have to be renormalized by a rapidity-width ratio, 
$\Gamma_y^M/\Gamma_y^\pi$ ($\Gamma_y^\pi$=1.8), 
to recover the proper equilibrium ratio for $dN_M/dy(y=0)$. 
In this way, our calculated meson spectra are absolutely normalized. 


Based on Eq.~(\ref{eqboltzappr}), the invariant meson $p_T$ spectrum 
takes the form
\begin{equation}
\label{ptspectra1}
E_M\frac{d N_M}{d^3 p}=\frac{d N_M}{d^2 p_T~dy}
=\frac{E_M}{(2\pi)^3}\int \frac{d^3x^*}{\gamma(x)}~ g(p,x)\gamma_p/\Gamma
\end{equation} 
with $g(p,x)$ defined in Eq.~(\ref{gainterm}).
In the azimuthally symmetric case, 3 integrations can be performed 
analytically, leaving a 3-dimensional integral to be computed numerically. 
With a dependence on the transverse radius of the form 
$(\frac{{\vec r}_T}{R_T})$, the eigenvolume of the fireball factorizes
and enters as an overall factor for $p_T$ spectra.
A more complete description of $p_T$ spectra should account for a hard 
component, which at RHIC energies is expected to become significant at 
$p_T\gtrsim 2 \div 3$~GeV. Since we here focus on the 
conceptual aspects of our approach, we will defer
the inclusion of a hard component to a future paper. 
 
Let us now specify the resonance parameters. 
The $\phi$ meson, which
in the vacuum has a mass and width of $m_{\phi}\simeq1020~\mbox{MeV}$ 
and $\Gamma_\phi \simeq4~\mbox{MeV}$, is expected to broaden 
substantially in hot and dense matter (cf., e.g., 
Ref.~\cite{vanHees:2006ng}), while mass shifts are more 
controversial. There are indications from lattice QCD that the
$\phi$ survives in the QGP at moderate temperatures 
above $T_c$~\cite{Asakawa:2002xj}. 
We will assume for the $\phi$ its vacuum mass and a
default width of $\Gamma_\phi=50$~MeV, in connection with a strange quark 
mass of $m_s=400$~MeV. 
The concept of $D$-meson resonances in the QGP has been implemented 
in Ref.~\cite{vanHees:2004gq} as a means to understand kinetic 
charm-quark equilibration at RHIC, with fair success in 
predicting~\cite{vanHees:2005wb} the most recent data for (decay-) 
electron suppression and $v_2$~\cite{Adare:2006nq}.  
Here we will assume $D$-mesons with a mass and 
width of $m_D=1.9$~{GeV} and $\Gamma_D=100$~MeV, and  
$c$- and $u$-quark masses of $m_c=1.5$~GeV and $m_u=350$~MeV.
For the $J/\psi$ we employ $m_{J/\psi}=3.1$~GeV, $\Gamma_{J/\psi}=100$~MeV.

\begin{figure}[t]
\includegraphics[width=0.55\textwidth]{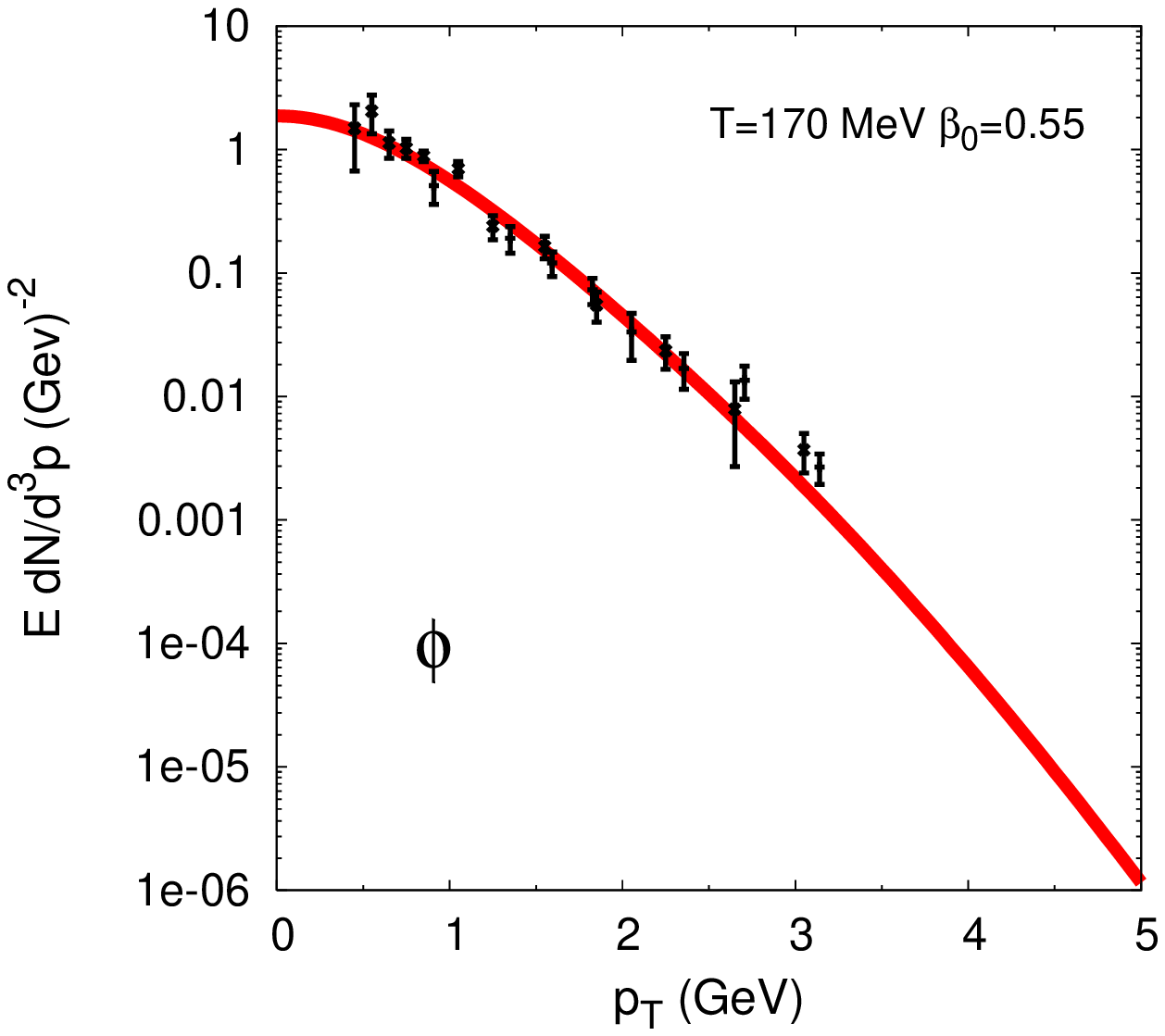}
\hspace{-2cm}
\includegraphics[width=0.55\textwidth]{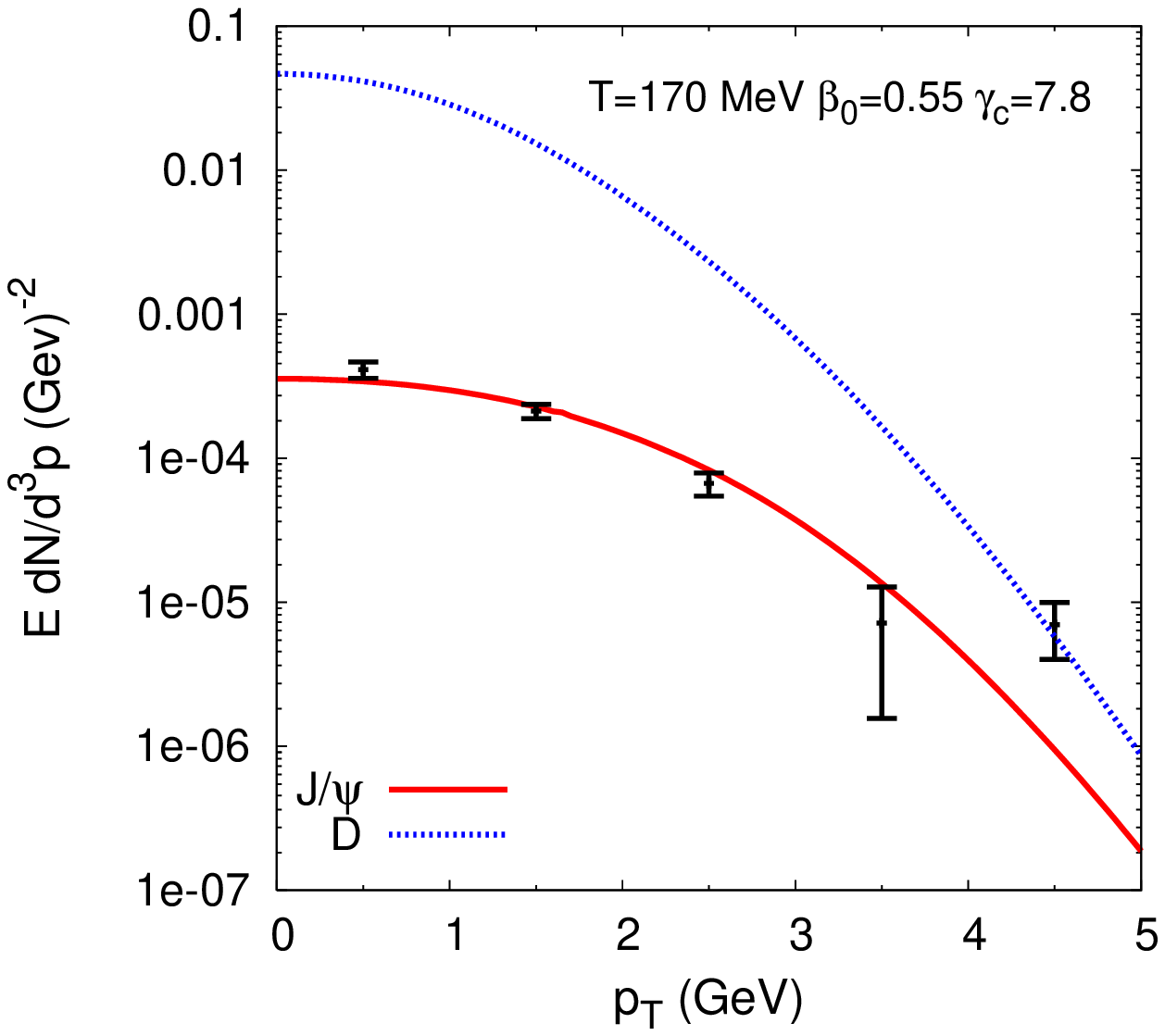} 
\begin{center}
\caption{Left panel: $p_T$ spectra for $\phi$ resonance formation
at $T_c=170$~MeV within our formalism, compared to data 
in central Au-Au at RHIC~\cite{Adler:2004hv,Adams:2004ux,Blyth:2006re}.  
Right panel: $p_T$ spectra for $J/\Psi$ (lower curve) and $D$ (upper curve)
resonance formation, compared to PHENIX $J/\Psi$ data
in central Au-Au~\cite{Adare:2006ns}; $\gamma_c(T=170~\mbox{MeV})=7.8$ 
has been taken from Ref.~\cite{Grandchamp:2003uw}.} 
\label{fig_pt}
\end{center}
\end{figure}

Fig.~\ref{fig_pt} illustrates our results for $\phi$ (left panel), $D$ and
$J/\psi$ (right panel) $p_T$ spectra compared to data in 
central $\sqrt{s_{NN}}=200$~GeV Au-Au. The surface flow value of 
$\beta_0=0.55$ has been adjusted to best reproduce
the $\phi$ $p_T$ spectra; it is comparable to the value used in the
conventional QCMs~\cite{Greco:2003mm}. The agreement is fair up to
the highest currently measured $p_T\simeq 3$~GeV. The  spectra are rather 
insensitive to variations in $\Gamma$ and $Q$ (provided that
$\Gamma\lesssim~Q$), but this will be different for the elliptic flow 
discussed below.
For $J/\psi$ and $D$ spectra we have introduced a charm-quark fugacity,
$\gamma_c$, which is necessary to match the total $c$-quark number
to the one expected from hard production in primordial nucleon-nucleon
collisions~\cite{pbm01,Grandchamp:2001pf}. Note that 
$N_{J/\psi}\propto \gamma_c^2$, with $\gamma_c$ subject
to appreciable uncertainty in the charm cross section as well as 
due to in-medium
effects on open-charm states~\cite{Grandchamp:2003uw}, 
cf.~also Ref.~\cite{Andronic:2006ky}. 

\section{Elliptic Flow and Scaling Properties}
\label{sec_elliptic}
Azimuthal asymmetries in the particle's momentum
distributions are defined via Fourier moments of order $n$,  
\begin{equation}
\label{v2anglephi}
v_n(p_T)=\langle\cos(n\varphi)\rangle(p_T) \ ,
\end{equation}
where the average is over $E dN/d^3p$ with $d^3p/E = dy dp_T^2 d\varphi$. 
At midrapidity, the odd moments vanish; in the following we will focus
on the elliptic flow, $v_2(p_T)$.  As discussed in the introduction, 
the experimental finding of an approximate CQNS relation for 
essentially all light hadrons at intermediate $p_T$,  
\begin{equation}
\label{v2scaling}
v_{2}^h(p_T)=n_Q~v_2^q(p_T/n_Q)
\end{equation}
is naturally reproduced by QCMs and constitutes one of their  
successes ($n_Q$: number of constituent quarks in hadron $h$).  
As in QCMs, our objective is to study how a given input $v_2^q$
in the quark distributions manifests itself on the hadron level.  
We adopt a factorized ansatz for the quark $v_2^q$, as used 
in previous coalescence studies~\cite{Greco:2003mm,Fries:2003kq}, 
\begin{equation}
\label{fqv2}
f_{q}(p,x,\varphi)=f_q(p,x)\otimes(1+2~ v_2^q(p_T)~\mbox{cos}(2\varphi))
\end{equation}
with $f_q(p,x)$ the distribution function of Eq.~(\ref{fq}).  
Note that this ``local" implementation of $v_2$ (cell by cell in the 
fireball) neglects space-momentum correlations characteristic for 
hydrodynamic expansion  
(more realistic distributions will be studied in forthcoming work). 
For the function $v_2^q(p_T)$ we employ a simplistic ansatz, which,
nevertheless, encodes the most important phenomenological features.
These are an essentially linear increase at low $p_T$ (as found in 
hydrodynamic 
models~\cite{Huovinen:2001cy,Teaney:2001av,Hirano:2002ds,Kolb:2003dz}), 
and a saturation at intermediate $p_T$, requiring two parameters: 
the transition momentum, and the saturation value of $v_2$, cf.
also Refs.~\cite{Greco:2003vf,Dong:2004ve}. For light 
quarks ($u$, $d$, $s$) we choose $p_T^{\rm sat}=1.1$~GeV and 
$v_2^{\rm sat}=7\%$, 
respectively, which gives a reasonable schematic representation
of the experimentally found scaling function~\cite{Adare:2006ti}. 
Similar features are borne out of parton cascade 
models~\cite{Molnar:2001ux,Zhang:2005ni}. For charm quarks, we take 
guidance from Langevin simulations for heavy quarks in a thermal 
background~\cite{vanHees:2005wb,Moore:2004tg}. Also here, a plateau
value of $v_2^{\rm sat}=7\%$ is within the uncertainty of current
data, while the transition momentum is
shifted to a higher value, $p_T^{\rm sat}\simeq2$~GeV, which 
is compatible with the mass ordering effect in hydrodynamic
calculations~\cite{Huovinen:2001cy}. We have verified that our 
schematic parametrization is consistent with a {\em parton-level} 
$KE_T$ scaling for $s$ and $c$ quarks (characteristic for a 
hydrodynamically expanding parton phase), where $KE_T=m_T-m$ denotes
the transverse kinetic energy of a particle ($m_T=\sqrt{m^2+p_T^2}$). 
$v_2$ saturation, implying deviation from hydrodynamics, 
signals, of course, incomplete thermalization and the transition
to a kinetic regime. 
\begin{figure}[t]
\includegraphics[width=0.55\textwidth]{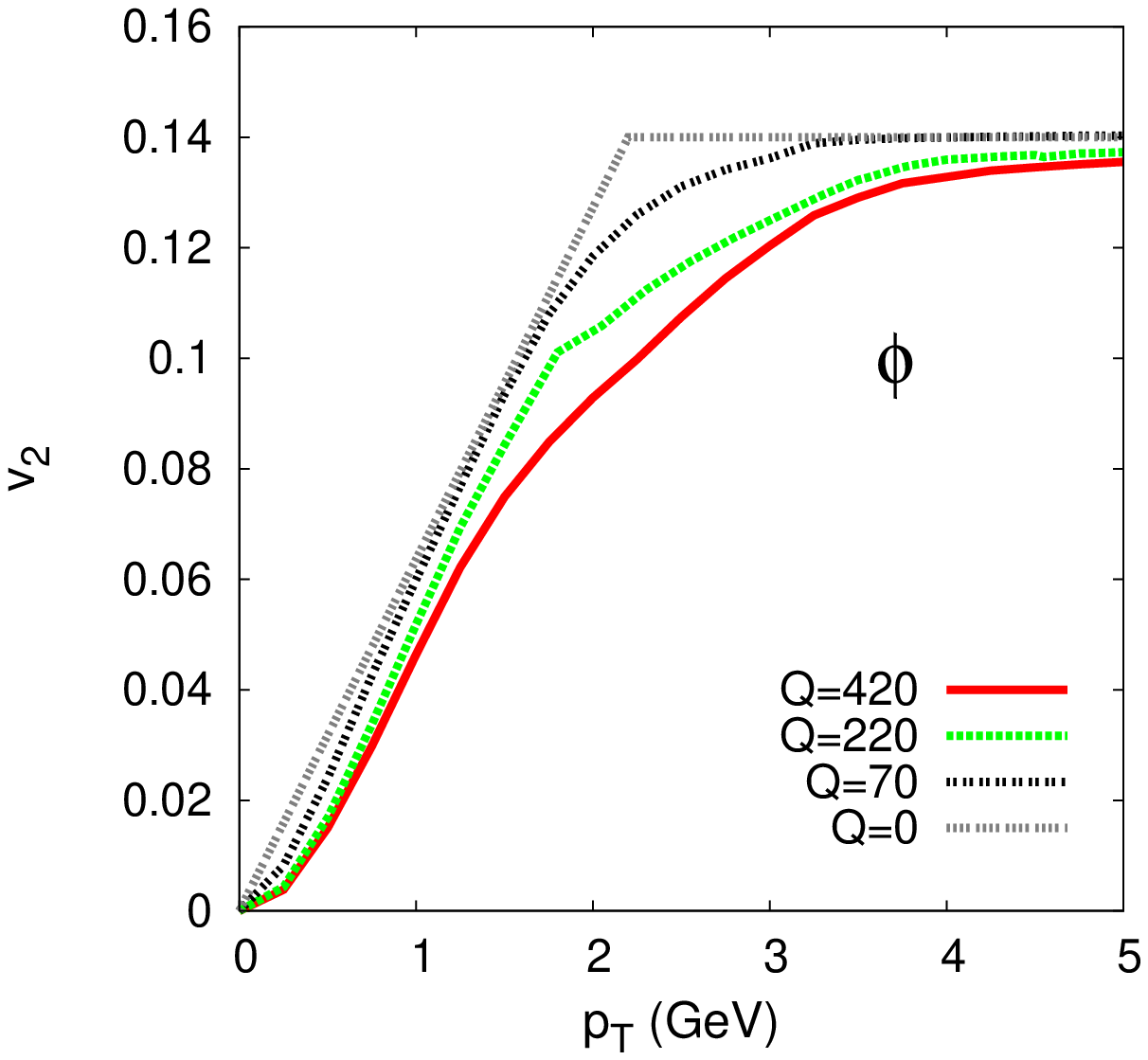}
\hspace{-2cm}
\includegraphics[width=0.55\textwidth]{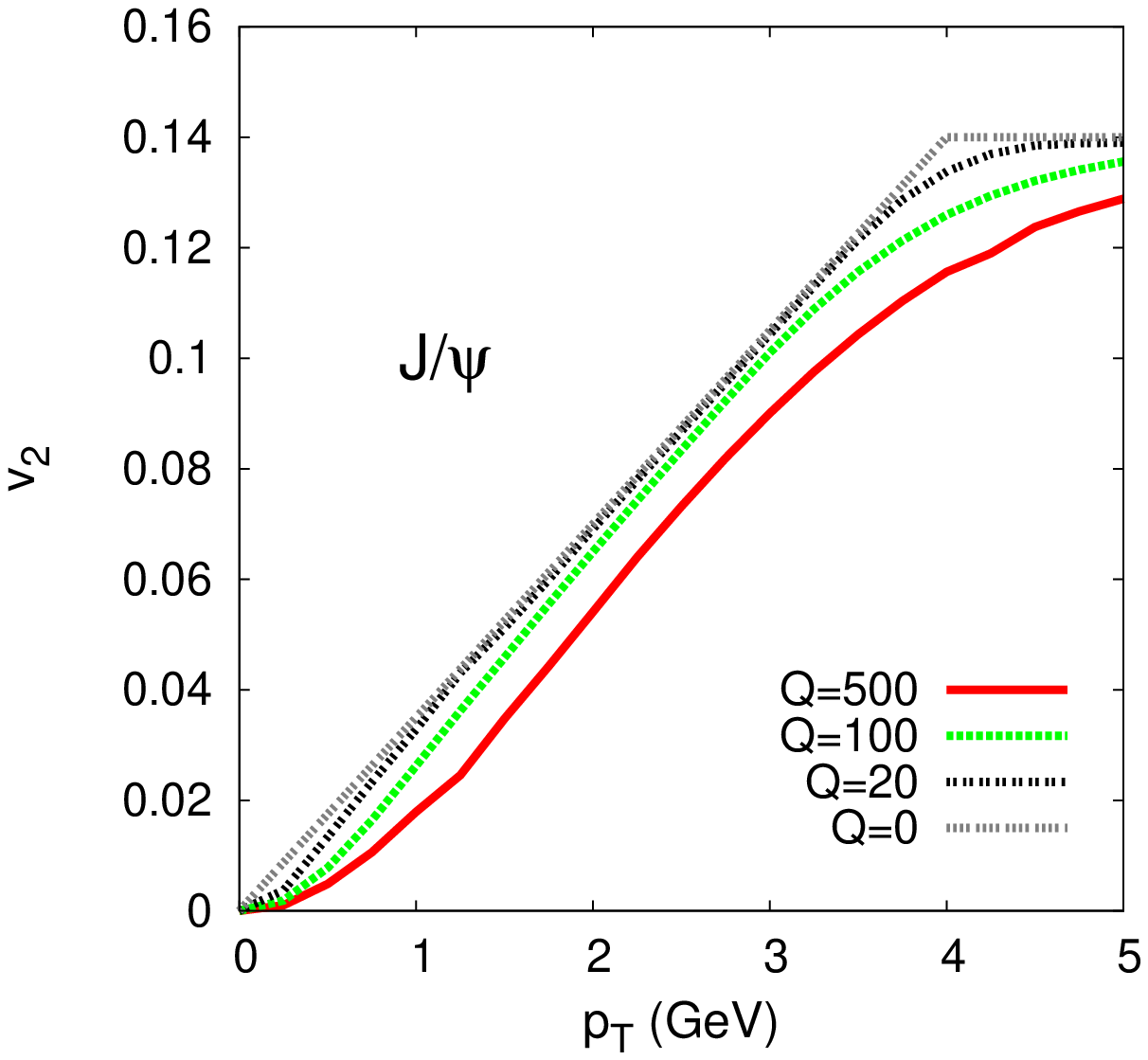}
\caption{Elliptic flow for $\phi$ (left panel) and
$J/\Psi$ (right panel) for 4 $Q$ values, 
$Q_\phi=420,220,70,\sim0~\mbox{MeV}$ and 
$Q_{J/\Psi}=500,100,20,\sim0~\mbox{MeV}$, from bottom up. 
The $Q\to0$ limits (upper curves) 
accurately reproduce the scaled quark $v_2$.} 
\label{fig_v2-1}
\end{figure}
In Fig.~\ref{fig_v2-1} we summarize our results for the $v_2(p_T)$ 
of $\phi$ and $J/\Psi$ mesons. The dependence of $v_2^M$ on $\Gamma$ is 
very weak, changing by less than 5\% when varying the $\phi$ width 
over the range $\Gamma=20-400$~MeV.
The dependence of $v_2^M$ on the $Q$ value for the resonances
is more pronounced as indicated by the different curves which have
been obtained by varying the underlying quark masses
(e.g., for the $\phi$, $Q=220$~MeV implies
$m_s=(m_\phi-Q)/2$=400~MeV). 
For any value of $Q$, CQNS, Eq.~(\ref{v2scaling}),
is recovered in our approach at sufficiently high $p_T$ where the 
production mechanism for an on-shell meson requires two constituent 
quarks with essentially collinear momenta, $p_q\sim p_{\bar{q}}\sim p_T/2$.
Consequently, the meson will inherit the full azimuthal asymmetry 
imparted by its constituents. As is well known, for collinear production
with $v_2^q\ll 1$, using Eq.~(\ref{fqv2}), the product $f_q f_{\bar q}$ 
recovers the scaling relation, Eq.~(\ref{v2scaling}).

Deviations from collinearity are expected to induce correction terms
involving $Q/p_T$. Since 4-momentum is conserved in our approach these 
corrections can be quantified. 
In particular at low $p_T$ there is no kinematical constraint 
enforcing collinearity of the reaction. For
$Q\to0$, we accurately reproduce CQNS at all $p_T$, with the meson 
$v_2$ exhibiting twice the asymptotic value of $v_2^q$ reached at 
exactly twice the quark transition momentum. 
For increasing $Q$, however, the convergence to the limiting value
is delayed to higher $p_T$, together with a reduction of
the $v_2$ at low $p_T$ (resembling the effect of a larger mass). 
\begin{figure}[t]
\includegraphics[width=0.55\textwidth]{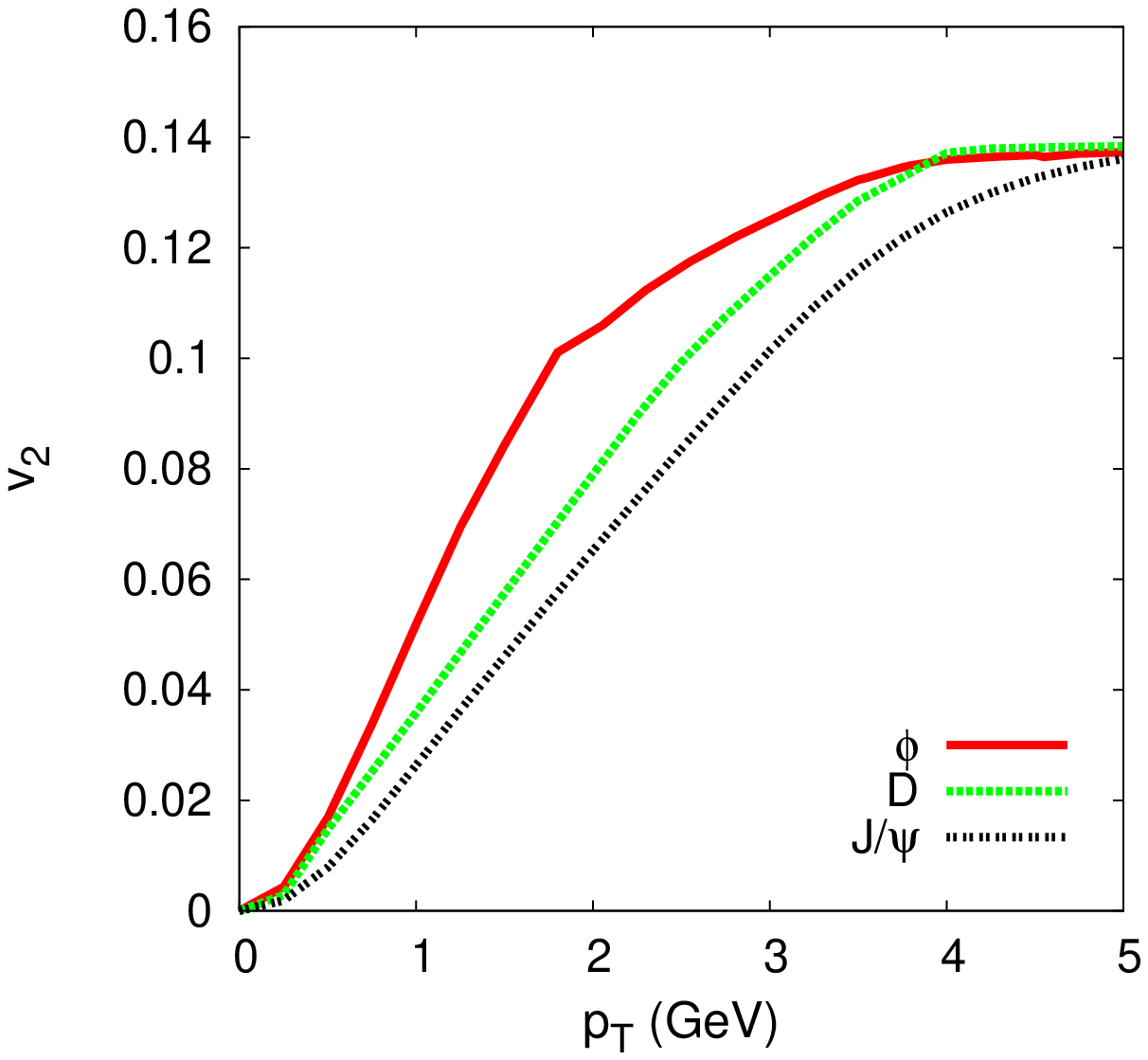}
\hspace{-2cm}
\includegraphics[width=0.55\textwidth]{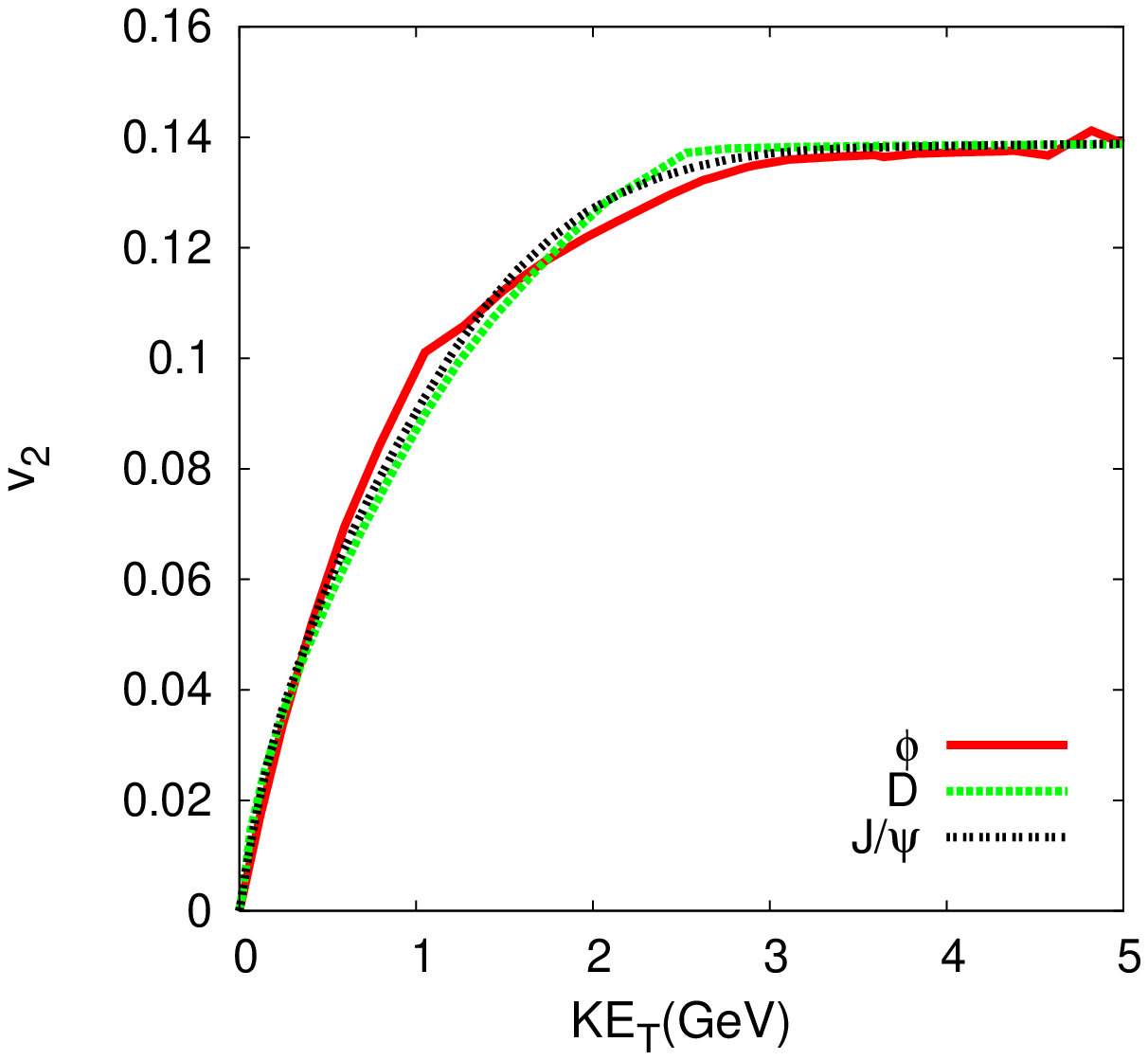}
\caption{Left panel: $v_2$ for $\phi$, $D$ and $J/\Psi$ (top to bottom)
using the default values for $Q$ and $\Gamma$; universal CQNS emerges 
only at high $p_T$. Right panel: $v_2$ for $\phi$, $D$ and $J/\Psi$ 
vs. kinetic energy, $KE_T$, of the meson; a universal scaling 
appears to emerge.}
\label{fig_v2-2}
\end{figure}

We finally return to the question of a universal CQNS.
A recent analysis of the PHENIX collaboration~\cite{Adare:2006ti} 
confirmed that CQNS for hadron $p_T$ spectra is only satisfied
at sufficiently high $p_T$, but the data exhibit a remarkably universal
CQNS scaling when plotted vs. kinetic energy of the hadrons, 
$KE_T=m_T-m$. At low $p_T$, $KE_T$ scaling is suggestive for hydrodynamic
behavior~\cite{Csanad:2005gv}, followed by a transition to a kinetic 
regime where the $v_2$ levels off.     
In Fig.~\ref{fig_v2-2} we summarize the CQNS properties as computed
in our approach. The left panel displays the $v_2$ for 
$\phi$, $D$, and $J/\Psi$ when scaled in $p_T$. The 3 curves are 
different at low $p_T$, converging at momenta as high as 
$p_T\gtrsim 5$~GeV. However, when the scaling is applied in  
$KE_T$ (right panel of Fig.~\ref{fig_v2-2}), the curves essentially
coincide over the entire energy range, in qualitative agreement
with experiment. While suggestive for the transition from the
hydrodynamic to the kinetic regime, it remains to be scrutinized 
in how far this result depends on the uncertainties
in the underlying parametrization of the quark $v_2$'s. 
As discussed above, the latter have been approximated by hydro-like
behavior at low $p_T$ with a sharp transition to an asymptotic value of 7\%
at a momentum as estimated from transport calculations and data.  
An interesting point is that a $KE_T$ scaling implemented 
at the partonic level appears to manifest itself as $KE_T$ scaling
at the meson level {\em if} the $Q$ value in the $q+\bar q \to M$
reaction is not too large (typically below 300~MeV, recall 
Fig.~\ref{fig_v2-1}). 
Another critical assumption is equality of the saturation value
for $v_2^q$ for $q=u,s,c$, which, in turn, is suggestive for full 
$c$-quark collectivity in a strongly coupled QGP (sQGP).      
We finally note that, especially for the $J/\psi$, the contribution 
from regeneration processes (as calculated here) 
is not expected to account for the yield at $p_T\gtrsim4$~GeV 
(cf.~right panel of Fig.~\ref{fig_pt}), which implies deviations
from CQNS at high momenta. 


\section{Conclusions}
We have proposed a reformulation of quark coalescence approaches   
based on a transport equation by implementing resonant quark-antiquark 
interactions to form mesons as the key microscopic mechanism.  
Our approach improves previous coalescence models by including 
explicit energy conservation in the meson production process and 
a well-defined thermal equilibrium limit. 
In the present exploratory study we have focused on a quasi-stationary
scenario by considering the meson formation process in a mixed 
phase at constant temperature, $T\simeq T_c$. 
We recover the empirically found constituent quark-number scaling 
(CQNS) in the meson $v_2$ at high momenta, while the improvements 
allow to extend the description to the low-momentum regime.
We have quantified deviations from CQNS in $p_T$ in terms of underlying 
spectral properties of the meson resonances, i.e., their widths and 
$Q$ values. Larger values of the latter are found to shift the
onset of CQNS to higher momenta. An interesting result of our study is
that, for input ($c$- and $s$-) quark distributions satisfying 
kinetic-energy scaling, a $KE_T$ scaling is recovered at the 
meson level ($\phi$, $D$, $J/\psi$) provided the $Q$-values are positive 
and not too large. A universal $KE_T$ scaling of hadron $v_2$ has 
recently been established experimentally~\cite{Adare:2006ti}.
To consolidate our results, the sensitivity to the input $v_2(p_T)$ at 
the quark level (including space-momentum correlations) needs to
be scrutinized~\cite{Pratt:2004zq}. 
Future work should also address the problem of baryon formation (e.g.,
within a quark-diquark picture~\cite{Miao:2007cm}), an explicit time
evolution (before and after $T_c$) as well as hadron channels with 
negative $Q$ values (e.g., for Goldstone bosons $\pi$ and $K$). 
While our framework does not address the hadronization
problem of gluons, we believe that it could lead to useful insights into 
systematic features of hadron production over a rather wide range of 
momenta.  

\begin{acknowledgments}
We thank Daniel Cabrera, Rainer Fries, Hendrik van Hees and 
Mauro Riccardi for valuable discussions.
This work was supported in part by a
U.S. National Science Foundation CAREER award under grant PHY-0449489.
\end{acknowledgments}

\end{document}